# Variability Aware Network Utility Maximization

Vinay Joseph and Gustavo de Veciana

Department of Electrical and Computer Engineering, The University of Texas at Austin

*Abstract*—**Network Utility Maximization (NUM) provides the key conceptual framework to study resource allocation amongst a collection of users/entities across disciplines as diverse as economics, law and engineering. In network engineering, this framework has been particularly insightfull towards understanding how Internet protocols allocate bandwidth, and motivated diverse research on distributed mechanisms to maximize network utility while incorporating new relevant constraints, on energy/power, storage, stability, etc., for systems ranging from communication networks to the smart-grid. However when the available resources and/or users' utilities vary over time, a user's allocations will tend to vary, which in turn may have a detrimental impact on the users' utility or quality of experience.**

**This paper introduces a generalized NUM framework which explicitly incorporates the detrimental impact of temporal variability in a user's allocated rewards. It explicitly incorporates tradeoffs amongst the mean and variability in users' allocations. We propose an online algorithm to realize variance-sensitive NUM, which, under stationary ergodic assumptions, is shown to be asymptotically optimal, i.e., achieves a time-average equal to that of an offline algorithm with knowledge of the future variability in the system. This substantially extends work on NUM to an intersting class of relevant problems where users/entities are sensitive to temporal variability in their service or allocated rewards.**

## I. INTRODUCTION

Network Utility Maximization (NUM) provides the key conceptual framework to study (fair) resource allocation among a collection of users/entities across disciplines as diverse as economics, law and engineering. In network engineering this framework has recently served as a particularly insightfull setting in which to study (reverse engineer) how the Internet's congestion control protocols allocate bandwidth, how to devise schedulers for wireless systems with time varying channel capacities, and motivated the development of distributed mechanisms to maximize network utility in diverse settings including communication networks and the smart grid, while incorporating new relevant constraints, on energy, storage, power control, stability, etc. However when the available resources and/or users' utilities vary over time, allocations amongst users will tend to vary, which in turn may have a detrimental impact on the users' utility or perceived service quality.

Indeed temporal variability in utility, service, resources or associated prices are particularly problematic when humans are the eventual recipients of the allocations. Humans typically view temporal variability negatively, as sign of an unreliable service, network or market instability, or as a service which when viewed through human's cognitive and behavioral responses can, and will, translate to a degraded Quality of

Experience (QoE). For example temporal variability in video quality has been shown to lead to hysteresis effects in humans quality judgments can can substantially degrade a user's QoE. This in turn can lead users to make decisions, e.g., change provider, act upon perceived market instabilities, etc., which can have serious implications on buisineses and engineered systems, or economic markets.

This paper introduces a generalized NUM framework which explicitly incorporates the detrimental impact of temporal variability in a user's allocated rewards. We use the term rewards as a proxy representing the resulting utility of, or any other quantity associated with, allocations to users/entities in a system. Our goal is to explicitly tackle the task of incorporating tradeoffs amongst the mean and variability in users' rewards. Thus, for example, in a variance-sensitive NUM setting, it may make sense to reduce a user's mean reward so as to reduce its variability. As will be discussed in the sequel there are many ways in which temporal variations can be accounted for, and which, in fact, present distinct technical challenges. In this paper we shall take a simple elegant approach to the problem which serves to address systems where tradeoffs amongst the mean and variability over time need to be made rather than systems where the mean (or target) is known, or where the issue at hand is the cumulative variance at the end of a given (e.g., investment) period.

To better describe the characteristics of the problem we introduce some preliminary notation. We shall consider a network shared by a set $\mathcal{N}$ of users (or other entities) where $|\mathcal{N}| = N$ denotes the number of users in the system Throughout the paper, we distinguish between random variables (and random functions) and their realizations by using upper case letters for the former and lower case for the latter. We use bold letters to denote vectors, e.g., $\mathbf{a} = (a_i : i \in \mathcal{N})$. We let $(\mathbf{a})_{1:T}$ denote the finite length sequence $(\mathbf{a}(t) : 1 \leq t \leq T)$. For a function $U$ on $\mathbb{R}$, $U'$ denotes its derivative.

Thus if $r_i(t)$ represents the reward allocated to user $i$ at time $t$, then $\mathbf{r}(t) = (r_i(t) : i \in \mathcal{N})$ is the vector of rewards to users $\mathcal{N}$ at time $t$ and $(\mathbf{r})_{1:T}$ represents the rewards allocated over time $t = 1, \ldots, T$ slots to the same users. We assume that reward allocations are subject to time varying network constraints,

$$c_t(\mathbf{r}(t)) \leq 1 \quad \text{for} \quad t = 1, \ldots, T,$$

where $c_t : \mathbb{R}^N \to \mathbb{R}$ corresponds to convex function, thus implicitly defining a convex set of feasible reward allocations. To formally capture the impact of the time-varying resources on users' QoE consider the following *offline* convex optimization

This research was supported in part by Intel and Cisco under the VAWN program, and by the NSF under Grant CNS-0917067.

problem OPT(T):

$$\max_{(\mathbf{r})_{1:T}} \sum_{i \in \mathcal{N}} \left( \underbrace{\frac{1}{T} \sum_{t=1}^{T} U_i^R \left( r_i(t) \right) - U_i^V \left( \text{Var}^T \left( (r_i)_{1:T} \right) \right)}_{\text{Proxy for user } i\text{'s QoE}} \right),$$

subject to 
$$c_t(\mathbf{r}(t)) \leq 0 \ \ \forall \ t \in \{1, ..., T\},$$
$$r_i(t) \geq r_{\min} \forall \ t \in \{1, ..., T\}, \forall \ i \in \mathcal{N},$$

where for each $i \in \mathcal{N}$,

$$\text{Var}^T \left( (r_i)_{1:T} \right) = \frac{1}{T} \sum_{t=1}^{T} \left( r_i(t) - \frac{1}{T} \sum_{\tau=1}^{T} r_i(\tau) \right)^2.$$

We refer to this as an offline optimization because time-varying time constraints $(c_t)_{1:T}$ are assumed to be known, and allow functions $\left( U_i^R, U_i^V \right)_{i \in \mathcal{N}}$ making the optimization problem convex. Note that the first term in a user $i$'s proxy QoE $\frac{1}{T} \sum_{t=1}^{T} U_i^R \left( r_i(t) \right)$ captures the degree to which QoE increases in his/her allocated rewards at any time, whereas the second term typically increasing in $\text{Var}^T(.)$ would penalizes temporal variability in reward allocation. Hence, this general formulation allows us to tradeoff between mean and variability associated with the reward allocations by appropriately choosing the functions $\left( U_i^R, U_i^V \right)_{i \in \mathcal{N}}$.

### A. Main result and contributions

The main contribution of this paper is in devising an *online* algorithm, for Adaptive Variability-Aware Resource (AVR) allocation, which realizes variance-sensitive NUM. Under stationary ergodic assumptions on the time-varying constraints, we show AVR is asymptotically optimal, i.e., achieves a performance equal to that of the offline optimization OPT(T) introduced earlier. This is a strong optimality result, which at first sight may be surprising due to the dependency of $\text{Var}^T(.)$ in the objective of OPT(T) on reward allocations over time and the time varying nature of the constraints $(c_t)_t$. The key idea exploits the characteristics of the problem, by keeping online estimates for the relevant quantities associated with users' allocations, e.g., the mean, variance, and mean QoE, which over time are shown to converge, and which eventually enable the online policy to produce allocations corresponding to the optimal stationary policy. Proving this result is somewhat challenging as it requires, showing that the estimates based on allocations produced by our online policy, AVR, (which itself depends on the estimated quantities), will converge to the desired values. To our knowledge this is the first attempt to generalize the NUM framework in this direction. We will contrast our problem formulation and approach to some of the past work in the literature addressing variance minimization, risk-sensitive control and other MDP based frameworks the related work below.

### B. Related Work

Network Utility Maximization (NUM) provides the key conceptual framework to study how to allocate rewards fairly amongst a collection of users/entities. [?] provides an overview of NUM. But all the work on NUM including several major extensions (for e.g., [?], [?], [?] etc.) have ignored the impact of variability in reward allocation on the quality of experience of users.

Adding a variance term in the objective function, would take things out of the general dynamic programming setting, see e.g. [?]. Indeed, including variance in the utility/cost to users at each time, signifies the overall cost is not decomposable, i.e., can not be written as a sum of costs each dependent only on the allocation at that time this makes sensitivity to variability challenging. For instance, [?] discusses minimum variance controller for linear systems (Section 5.3) where the objective is the minimization of the sum of second moments of the output variable. Sum of second moments is considered instead of the variance, which allows the cumulative cost to be represented sum of the costs incurred over time. Note however, that minimization of second moments does not directly address variability unless the mean is zero. The variance of the cumulative cost is incorporated in the objective for problems in risk sensitive optimal control (see [?]) to capture the risk associated with a policy. Note however, that the variance is of the cumulative cost rather than of the variability as seen by a user over time. To summarize, to our knowledge there are no previously proposed works on NUM that addresses the negative impact of variability. The algorithm proposed here falls into the class of stochastic fixed point algorithms (see [?]). Our algorithm is also related to the algorithms proposed in [?] and [?] although these works also ignore variability.

### C. Organization of the paper

In Section II, we discuss the system model and assumptions. We study the optimality conditions for OPT(T) in Section III. We introduce OPTSTAT in IV and study its optimality conditions. We start Section V by formally introducing our online algorithm AVR. Then do a convergence analysis of AVR in Subsection V-A, and conclude the section by establishing the asymptotic optimality of AVR in Subsection V-B. We conclude the paper in Section VI. The proofs of some of the intermediate results used in the paper are discussed in an appendix given at the end of the paper.

## II. SYSTEM MODEL

We consider a slotted system where slots are indexed by $t \in \{0, 1, 2...\}$, and the system serves a fixed set of users $\mathcal{N}$ and let $N = |\mathcal{N}|$.

Let $\mathbb{R}_+ = \{b \in \mathbb{R} : b \geq 0\}$. A sequence $(\mathbf{b}(t))_t$ in a Euclidean space is said to converge to a *set* $\mathcal{A}$ if

$$\lim_{t \to \infty} \inf_{\mathbf{a} \in \mathcal{A}} \|\mathbf{b}(t) - \mathbf{a}\| = 0,$$

where $\|.\|$ denotes the Euclidean norm associated with the space. For a function $U$ on $\mathbb{R}$, $U^{'}$ denotes its derivative. We use $\mathbf{I}$ as the indicator function, i.e., for any set $\mathcal{A}$, we let $\mathbf{I}_{\{a \in \mathcal{A}\}} = 1$ if $a \in \mathcal{A}$, and zero otherwise.

We assume that the reward allocation $\mathbf{r}(t) \in \mathbb{R}_+^N$ in slot $t$ is constrained to satisfy the following inequality

$$c_t(\mathbf{r}(t)) \leq 0,$$

where $c_t$ is picked from a (arbitrarily large) finite set $\mathcal{C}$ of real valued maps on $\mathbb{R}_+^N$. We make the following assumptions on these constraints:

---

**Assumptions C1-C4 (Time varying constraints)**

**C.1** There is a constant $r_{\min} \geq 0$ such that for any $c \in \mathcal{C}$, $c(\mathbf{r}) \leq 0$ for $\mathbf{r}$ such that $r_i = r_{\min}$ for each $i \in \mathcal{N}$.

**C.2** There is a constant $r_{\max} > 0$ such that for any $c \in \mathcal{C}$ and $\mathbf{r} \in \mathbb{R}_+^N$ satisfying $c(\mathbf{r}) \leq 0$, we have $r_i \leq r_{\max}$ for each $i \in \mathcal{N}$.

**C.3** Each function $c \in \mathcal{C}$ is convex and differentiable on an open set containing $[r_{\min}, r_{\max}]$.

**C.4** For any $c \in \mathcal{C}$ and $\mathbf{r}$ such that $r_i = r_{\min}$ for each $i \in \mathcal{N}$, $c(\mathbf{r}) < 0$ or $c(\mathbf{r}) \leq 0$ if $c$ is an affine function.

**C.5** Let $(C_t)_t$ be a stationary ergodic process, and let $(\pi(c) : c \in \mathcal{C})$ denote the stationary distribution associated.

---

We let $C^\pi$ denote a random constraint with distribution $(\pi(c) : c \in \mathcal{C})$.

We could allow the constants $r_{\min}$ and $r_{\max}$ to be user dependent. But, we avoid that for notational simplicity. The condition C.4 is imposed to ensure that the constraint set is 'nice' when used as a feasible set for an optimization problem OPT(T) (see for e.g. Lemma 1).

Next we discuss the assumptions on the functions $\left( U_i^R, U_i^V \right)_{i \in \mathcal{N}}$. For each $i \in \mathcal{N}$, we make the assumptions U.V and U.R discussed next.

---

**Assumptions U.V and U.R**

Let $v_{\max} = (r_{\max} - r_{\min})^2$.

**U.V:** $U_i^V$ is defined and twice continuously differentiable on an open set containing $[0, v_{\max}]$ with $\min_{v \in [0, v_{\max}]} \left( U_i^V \right)'(v) = d_{\min,i}^V > 0$ and $\min_{v \in [0, v_{\max}]} \left( U_i^V \right)''(v) < 0$. Further, we assume that for any two elements $\mathbf{x}^1$ and $\mathbf{x}^2$ in any Euclidean space $\mathbb{R}^d$ with $\mathbf{x}^1 \neq \mathbf{x}^2$, and $\alpha \in (0,1)$ with $\bar{\alpha} = 1 - \alpha$, we have

$$U_i^V \left( \|\alpha \mathbf{x}^1 + \bar{\alpha} \mathbf{x}^2\|^2 \right) < \alpha U_i^V \left( \|\mathbf{x}^1\|^2 \right) + \bar{\alpha} U_i^V \left( \|\mathbf{x}^2\|^2 \right), \quad (1)$$

where $\|.\|$ denotes the Euclidean norm associated with the space.

For each $i \in \mathcal{N}$, let $\max_{v \in [0, v_{\max}]} \left( U_i^V \right)'(v) = d_{\max,i}^V$.

**U.R:** $U_i^R$ is defined and differentiable on an open set containing $[r_{\min}, r_{\max}]$. Further, we assume that $U_i^R$ is concave and strictly increasing on $[r_{\min}, r_{\max}]$.

---

Note that by picking, for each $i \in \mathcal{N}$, the functions $U_i^V$ from the following set

$$\mathcal{U}_V = \left\{ (v + \delta)^\alpha : \alpha \in [0.5, 1] \text{ with } \delta > 0 \text{ if } \alpha \neq 1 \right\},$$

we satisfy the requirements in U.V. Note that this includes the identity function $U^V(v) = v$. Also, the function $U^V(v) = \sqrt{v + \delta}$ for any (arbitrarily small) $\delta > 0$ satisfies the conditions in U.V.

We satisfy U.R if we pick the functions $\left( U_i^R \right)_{i \in \mathcal{N}}$ from following class of strictly concave increasing functions

parametrized by $\alpha \in (0, \infty)$ ([**?**])

$$U_\alpha(x) = \begin{cases} \log x & \text{if } \alpha = 1, \\ (1 - \alpha)^{-1} x^{1 - \alpha} & \text{otherwise,} \end{cases} \quad (2)$$

These functions are commonly used to enforce fairness to obtain allocations that are $\alpha-$fair (see [**?**]). A larger $\alpha$ corresponds to a more fair allocation. Note that we have to ensure that $0 \notin [r_{\min}, r_{\max}]$ to ensure that function is well defined, and even if this is not the case, we could use $U_\alpha(.+\delta)$ instead of $U_\alpha(.)$ for an arbitrarily small positive shift $\delta$ in the argument to avoid this requirement.

We will see later that AVR can be made more efficient if $U_i^V$ is linear for some users $i \in \mathcal{N}$. We define the following subsets of $\mathcal{N}$:

$$\mathcal{N}_{Vl} = \left\{ i \in \mathcal{N} : U_i^V \text{ is linear} \right\},$$
$$\mathcal{N}_{Vn} = \left\{ i \in \mathcal{N} : U_i^V \text{ is not linear} \right\}.$$

We focus on obtaining an algorithm for reward allocation that can be implemented at a centralized coordinator that has access to $c_t$ at the beginning of slot $t$. For instance, in a cellular network setting (like in WN), this could be a basestation that estimates the channel strengths of the users in the network to find $c_t$.

### A. Applications and scope of the model

The presence of time varying constraints $c_t(\mathbf{r}) \leq 0$ allows us to apply the model to several interesting and useful settings. In particular, here we focus on a wireless network setting by discussing three cases WN, WN-E and WN-T, and show that the model can handle problems involving time varying exogenous constraints and time varying utility functions. We start by discussing case WN where the reward in a slot is the rate allocated to the user in that slot. Let $\mathcal{P}$ denote a finite (but arbitrarily large) set of positive vectors where each vector corresponds to the peak transmission rate vector for a slot seen by users in a wireless network. Let $\mathcal{C} = \left\{ c_{\mathbf{p}} : c_{\mathbf{p}}(\mathbf{r}) = \sum_{i \in \mathcal{N}} \frac{r_i}{p_i} - 1, \ \mathbf{p} \in \mathcal{P} \right\}$. Here, for any allocation $\mathbf{r}$, $r_i/p_i$ is the fraction of time the wireless system needs to serve user $i$ in slot $t$ to deliver data at the rate of $r_i$ to user $i$ in a slot where the user has peak data transmission rate $p_i$. Thus, the constraint $c_{\mathbf{p}}(\mathbf{r}) \leq 0$ can be seen as a scheduling constraint that corresponds to the requirement that the sum of the fractions of time that different users are served in a slot should be less than or equal to one.

**Time varying exogenous constraints**: We can also allow for time varying exogenous constraints on the wireless system by appropriately defining the set $\mathcal{C}$. For instance, consider case WN-E where a base station in a cellular network allocates rates to users some of whom are streaming videos. As pointed above QoE of users viewing video content is sensitive to temporal variability in quality. But, while allocating rates to these users, we also need to account for the time varying resources requirements of the voice and data traffic handled by the basestation. We can deal with this constraint by defining $\mathcal{C} = \left\{ c_{(\mathbf{p},f)} : c_{(\mathbf{p},f)}(\mathbf{r}) = \sum_{i \in \mathcal{N}} \frac{r_i}{p_i} - (1 - f), \ \mathbf{p} \in \mathcal{P}, f \in \mathcal{F} \right\}$, where $\mathcal{F}$ is a finite set of real numbers in $[0, 1]$ where each element in the set corresponds to the fraction of time in a slot that is utilized the voice and data traffic.

**Time varying utility functions**: For the users streaming video content discussed in the case WN-E, it is more appropriate to view the perceived video quality of a user in a slot as the reward for that user in that slot. However, for users streaming video content, the dependence of perceived video quality (in a short duration slot roughly a second long which corresponds to a collection of 20-30 frames) on the compression rate is time varying. This is typically due to the possibly changing nature of the content, e.g., from an action to a slower scene. Hence, the 'utility' function that maps the reward (i.e., perceived video quality) derived from the allocated resource (i.e., the rate) is time varying. This is the setting in the case WN-T, and we can handle it as follows. Let $q_{t,i}(w_i)$ denote the strictly increasing concave function that, in slot $t$, maps the perceived video quality to the rate $w_i$ allocated to user $i$. For each user $i$, let $\mathcal{Q}_i$ be a finite set of such functions. Hence, we can view WN-T as a case that has the following set of constraints:

$$\mathcal{C} = \left\{ c_{(\mathbf{p},\mathbf{q})} : c_{(\mathbf{p},\mathbf{q})}(\mathbf{r}) = \sum_{i \in \mathcal{N}} \frac{q_i^{-1}(r_i)}{p_i} - 1, \right.$$
$$\left. \mathbf{p} \in \mathcal{P}, q_i \in \mathcal{Q}_i \ \forall \ i \in \mathcal{N} \right\}.$$

Note that each element in $\mathcal{C}_3$ is a convex function.

For WN and WN-E, we can verify that by choosing $r_{\max} = \max_{\mathbf{p} \in \mathcal{P}} \max_{i \in \mathcal{N}} p_i$ and an $r_{\min}$ satisfying $0 \leq r_{\min} \leq \frac{1}{N} \min_{\mathbf{p} \in \mathcal{P}} \min_{i \in \mathcal{N}} p_i$, we satisfy C.1-C.4. In WN-T, if we assume that each function $q \in \mathcal{Q}$ is differentiable and convex with $q(0) = 0$ (which are very reasonable assumptions on the dependence between quality and compression rate), then we can verify that by choosing $r_{\min} = 0$ and $r_{\max} = \max_{\mathbf{p} \in \mathcal{P}} \max_{i \in \mathcal{N}} \max_{q \in \mathcal{Q}} q(p_i)$, we satisfy C.1-C.4.

**Variability aware rate adaptation for video**: The above formulation is applicable to the problem of finding optimal (joint) video rate adaptation that maximizes the sum QoE of users streaming videos utilizing resources of a shared network. Given the predictions for explosive growth of video traffic in the near future (see [**?**]), this is among one of the important networking problems today. For a user viewing a video stream, variations in video quality over time has a detrimental impact on the user's QoE, see e.g., [**?**], [**?**], [**?**]. Indeed [**?**] even points out that variations in quality can result in a QoE that is worse than that of a constant quality video with lower average quality. Furthermore, [**?**] proposed and evaluated a metric for QoE which roughly corresponds to the choices $U_i^R(r) = r$ and $U_i^V(v) = \sqrt{v + \delta}$ in the model described above for a very small $\delta > 0$.

## III. OPTIMAL VARIANCE-SENSITIVE OFFLINE POLICY

In this section, we study OPT(T), the offline formulation for optimal joint reward allocation introduced in Section I. In the offline setting, we assume that $(c)_{1:T}$, i.e., the realization of the process $(C)_{1:T}$, is known. We denote the objective function of OPT(T) by $\phi_T$, i.e.,

$$\phi_T\left((\mathbf{r})_{1:T}\right)$$
$$= \sum_{i \in \mathcal{N}} \left( \frac{1}{T} \sum_{t=1}^{T} U_i^R(r_i(t)) - U_i^V\left(\text{Var}^T\left((r_i)_{1:T}\right)\right) \right),$$

and $\left(U_i^R\right)_{i \in \mathcal{N}}$ and $\left(U_i^V\right)_{i \in \mathcal{N}}$ are functions satisfying U.R and U.V respectively, and $\text{Var}^T\left((r_i)_{1:T}\right) = \frac{1}{T} \sum_{t=1}^{T} \left(r_i(t) - \frac{1}{T} \sum_{\tau=1}^{T} r_i(\tau)\right)^2$. Hence the optimization problem OPT($T$) can be rewritten as:

$$\max_{(\mathbf{r})_{1:T}} \phi_T\left((\mathbf{r})_{1:T}\right) \tag{3}$$
$$\text{subject to} \quad c_t(\mathbf{r}(t)) \leq 0 \ \forall \ t \in \{1, ..., T\}, \tag{4}$$
$$r_i(t) \geq r_{\min} \forall \ t \in \{1, ..., T\}, \forall \ i \in \mathcal{N}, \tag{5}$$

where $c_t \in \mathcal{C}$ is a convex function for each $t$. The next result asserts that OPT($T$) is a convex optimization problem satisfying Slater's condition (Section 5.2.3, [**?**]) and that it has a unique solution.

**Lemma 1.** *OPT($T$) is a convex optimization problem satisfying Slater's condition with a unique solution.*

*Proof:* Since we made the assumptions U.R and U.V, the convexity of the objective of OPT($T$) is easy to establish once we prove the convexity of the function $U_i^V(\text{Var}^T(.))$ for each $i \in \mathcal{N}$. Using (1) and the definition of $\text{Var}^T(.)$, we can show that $U_i^V(\text{Var}^T(.))$ is a convex function for each $i \in \mathcal{N}$. The details are given next. For two different quality vectors $(\mathbf{r^1})_{1:T}$ and $(\mathbf{r^2})_{1:T}$, any $i \in \mathcal{N}$, $\alpha \in (0, 1)$ and $\bar{\alpha} = 1 - \alpha$, we have that

$$\text{Var}^T\left(\alpha\left(r_i^1\right)_{1:T} + \bar{\alpha}\left(r_i^2\right)_{1:T}\right)$$
$$= \text{Var}^T\left(\left(\alpha r_i^1 + \bar{\alpha} r_i^2\right)_{1:T}\right)$$
$$= \frac{1}{T} \sum_{t=1}^{T} \left(\left(\alpha r_i^1(t) + \bar{\alpha} r_i^2(t)\right)\right.$$
$$\left. - \frac{1}{T} \sum_{\tau=1}^{T} \left(\alpha r_i^1(\tau) + \bar{\alpha} r_i^2(\tau)\right)\right)^2$$
$$= \frac{1}{T} \sum_{t=1}^{T} \left(\alpha\left(r_i^1(t) - \frac{1}{T} \sum_{\tau=1}^{T} r_i^1(\tau)\right)\right.$$
$$\left. + \bar{\alpha}\left(r_i^2(t) - \frac{1}{T} \sum_{\tau=1}^{T} r_i^2(\tau)\right)\right)^2$$

Using (1), we have that

$$U_i^V\left(\text{Var}^T\left(\alpha\left(r_i^1\right)_{1:T} + \bar{\alpha}\left(r_i^2\right)_{1:T}\right)\right)$$
$$\leq \alpha U_i^V\left(\frac{1}{T} \sum_{t=1}^{T} \left(r_i^1(t) - \frac{1}{T} \sum_{\tau=1}^{T} r_i^1(\tau)\right)^2\right)$$
$$+ \bar{\alpha} U_i^V\left(\frac{1}{T} \sum_{t=1}^{T} \left(r_i^2(t) - \frac{1}{T} \sum_{\tau=1}^{T} r_i^2(\tau)\right)^2\right)$$
$$= \alpha U_i^V\left(\text{Var}^T\left((r_i^1)_{1:T}\right)\right) + \bar{\alpha} U_i^V\left(\text{Var}^T\left((r_i^2)_{1:T}\right)\right).$$

Thus, $U_i^V\left(\text{Var}^T(.)\right)$ is a convex function. Using the above arguments and concavity of $U_i^R$ and $-U_i^V\left(\text{Var}^T(.)\right)$, we conclude that OPT$(T)$ is a convex optimization problem.

Note that, from (1) (since we have a strict inequality), the inequality above is a strict one unless

$$r_i^1(t) = r_i^2(t) + \frac{1}{T}\sum_{\tau=1}^{T} r_i^1(\tau) - \frac{1}{T}\sum_{\tau=1}^{T} r_i^2(\tau) \ \forall \ 1 \le t \le T.$$

Thus, for the inequality not to be a strict one, we require that $\text{Var}^T\left(\left(r_i^1\right)_{1:T}\right) = \text{Var}^T\left(\left(r_i^2\right)_{1:T}\right)$. Further, Slater's condition is satisfied and it mainly follows from the assumption C.4.

Now, for any $i \in \mathcal{N}$, $U_i^R$ and $-U_i^V\left(\text{Var}^T(.)\right)$ are not necessarily strictly concave. But, we can still show that the objective is strictly concave as follows. Let $\left(\mathbf{r^1}\right)_{1:T}$ and $\left(\mathbf{r^2}\right)_{1:T}$ be two optimal solutions to OPT$(T)$. Then, from the concavity of the objective, $\left(\alpha\left(r_i^1\right)_{1:T} + \bar{\alpha}\left(r_i^2\right)_{1:T}\right)$ is also an optimal solution for any $\alpha \in (0,1)$ and $\bar{\alpha} = 1 - \alpha$. Due to concavity of $U_i^R(.)$ and convexity of $U_i^V\left(\text{Var}^T(.)\right)$, this is only possible if for each $i \in \mathcal{N}$ and $1 \le t \le T$, $U_i\left(\alpha r_i^1(t) + \bar{\alpha} r_i^2(t)\right) = \alpha U_i^R\left(r_i^1(t)\right) + \bar{\alpha}U R_i\left(r_i^2(t)\right)$, and $U_i^V\left(\text{Var}^T\left(\alpha\left(r_i^1\right)_{1:T} + \bar{\alpha}\left(r_i^2\right)_{1:T}\right)\right) = \alpha U_i^V\left(\text{Var}^T\left(\left(r_i^1\right)_{1:T}\right)\right) + \bar{\alpha}U_i^V\left(\text{Var}^T\left(\left(r_i^2\right)_{1:T}\right)\right)$.

From above discussion, $U_i^V\left(\text{Var}^T\left(\alpha\left(r_i^1\right)_{1:T} + \bar{\alpha}\left(r_i^2\right)_{1:T}\right)\right)$ is equal to $\alpha U_i^V\left(\text{Var}^T\left(\left(r_i^1\right)_{1:T}\right)\right) + \bar{\alpha}U_i^V\left(\text{Var}^T\left(\left(r_i^2\right)_{1:T}\right)\right)$ for each $i \in \mathcal{N}$ only if $\text{Var}^T\left(\left(r_i^1\right)_{1:T}\right) = \text{Var}^T\left(\left(r_i^2\right)_{1:T}\right)$ for each $i \in \mathcal{N}$, and $r_i^1(t) = r_i^2(t) + \frac{1}{T}\sum_{\tau=1}^{T} r_i^1(\tau) - \frac{1}{T}\sum_{\tau=1}^{T} r_i^2(\tau)$ for each $i \in \mathcal{N}$ and $1 \le t \le T$. Since for each $i \in \mathcal{N}$, $\text{Var}^T\left(\left(r_i^1\right)_{1:T}\right) = \text{Var}^T\left(\left(r_i^2\right)_{1:T}\right)$, due to optimality of $\left(\mathbf{r^1}\right)_{1:T}$ and $\left(\mathbf{r^2}\right)_{1:T}$, we have that

$$\sum_{i \in \mathcal{N}}\left(\frac{1}{T}\sum_{t=1}^{T} U_i^R\left(r_i^2(t)\right) - U_i^V\left(\text{Var}^T\left(\left(r_i^2\right)_{1:T}\right)\right)\right)$$

$$= \sum_{i \in \mathcal{N}}\left(\frac{1}{T}\sum_{t=1}^{T} U_i^R\left(r_i^1(t)\right) - U_i^V\left(\text{Var}^T\left(\left(r_i^2\right)_{1:T}\right)\right)\right)$$

$$= \sum_{i \in \mathcal{N}}\left(\frac{1}{T}\sum_{t=1}^{T} U_i^R\left(r_i^2(t) \right.\right.$$

$$\left.\left. + \frac{1}{T}\sum_{\tau=1}^{T} r_i^1(\tau) - \frac{1}{T}\sum_{\tau=1}^{T} r_i^2(\tau)\right) - U_i^V\left(\text{Var}^T\left(\left(r_i^2\right)_{1:T}\right)\right)\right)$$

Since $U_i^R$ is a strictly increasing function for each $i \in \mathcal{N}$, the above equation implies that

$$\frac{1}{T}\sum_{\tau=1}^{T} r_i^1(\tau) = \frac{1}{T}\sum_{\tau=1}^{T} r_i^2(\tau),$$

and thus,

$$r_i^1(t) = r_i^2(t) \ \forall \ 1 \le t \le T, \ \forall \ i \in \mathcal{N}.$$

From the above discussion, we can conclude that OPT$(T)$ has a unique solution. ∎

We let $\left(\mathbf{r}^T\right)_{1:T}$ denote the optimal solution to OPT$(T)$. Since OPT$(T)$ is a convex optimization problem satisfying Slater's condition (Lemma 1), Karush-Kuhn-Tucker (KKT) conditions ([?]) given next are necessary and sufficient for optimality. Let $m_i^T = \frac{1}{T}\sum_{t=1}^{T} r_i^T(t)$.

---

**KKT-OPT(T):**

$\left(\mathbf{r}^T\right)_{1:T}$ is an optimal solution to OPT$(T)$ if and only if it is feasible, and there exist non-negative constants $\left(\mu^T\right)_{1:T}$ and $\left(\gamma_i^T : i \in \mathcal{N}\right)_{1:T}$ such that for all $i \in \mathcal{N}$ and $t \in \{1, ..., T\}$, we have

$$\frac{\left(U_i^R\right)'\left(r_i^T(t)\right)}{T} - \frac{2\left(U_i^V\right)'\left(\text{Var}\left(\left(r_i^T\right)_{1:T}\right)\right)}{T}\left(r_i^T(t) - m_i^T\right)$$

$$- \frac{\mu^T(t)}{T}c_{t,i}'(\mathbf{r}^T(t)) + \frac{\gamma_i^T(t)}{T} = 0, \quad (6)$$

$$\mu^T(t)c_t(\mathbf{r}^T(t)) = 0, \quad (7)$$

$$\gamma_i^T(t)\left(r_i^T(t) - r_{\min}\right) = 0, \quad (8)$$

---

Here $c_{t,i}'$ denotes $\frac{\partial c_t}{\partial r_i}$, and we have used the fact that for any $i \in \mathcal{N}$ and $\tau' \in \{1, ..., T\}$

$$\frac{\partial}{\partial r_i(\tau')}\left(T\text{Var}^T\left((r_i)_{1:T}\right)\right) = 2\left(r_i(\tau') - \frac{1}{T}\sum_{\tau=1}^{T} r_i(\tau)\right).$$

From (6), we see that the optimal reward allocation $\mathbf{r}^T(t)$ in any time slot $t$ depends on the entire allocation $\left(\mathbf{r}^T\right)_{1:T}$ only through the following four quantities associated with $\left(\mathbf{r}^T\right)_{1:T}$: (i) time average reward $\mathbf{m}^T$, (ii) $\left(\left(U_i^V\right)'\right)_{i \in \mathcal{N}}$ evaluated at the variance seen by the respective users. So, if a genie revealed these quantities, the optimal allocation for each slot $t$, can be determined by solving an optimization that only requires the knowledge of $c_t$ (associated with current slot) and not $(c)_{1:T}$. We exploit this key idea while formulating the online algorithm AVR (proposed in Section V).

## IV. A RELATED PROBLEM: OPTSTAT

In this section, we introduce and study another optimization problem OPTSTAT closely related to OPT$(T)$. The formulation OPT(T) mainly involves time averages of various quantities associated with it. Instead, the formulation of OPTSTAT is based on the expected value of the corresponding quantities evaluated using the stationary distribution of $(C_t)_t$.

Recall that (see C.5) $(C_t)_t$ is a stationary ergodic process with stationary distribution $(\pi(c) : c \in \mathcal{C})$, i.e., for $c \in \mathcal{C}$, $\pi(c)$ is the probability of the event $c_t = c$. Since $\mathcal{C}$ is finite, we assume that $\pi(c) > 0$ for each $c \in \mathcal{C}$ without any loss of generality. Let $(\mathbf{r}(c))_{c \in \mathcal{C}}$ be a vector representing the reward allocation $\mathbf{r}(c)(\in \mathbb{R}^N)$ to the users for each $c \in \mathcal{C}$. Although we are abusing the notation introduced earlier where $\mathbf{r}(t)$ denoted the allocation to the users in slot $t$, one can differentiate between the functions based on the context in which they are being discussed. Now, let

$$\phi_\pi\left((\mathbf{r}(c))_{c \in \mathcal{C}}\right) = \sum_{i \in \mathcal{N}}\left(\sum_{c \in \mathcal{C}} \pi(c)U_i^R\left(r_i(c)\right)\right.$$

$$\left. - U_i^V\left(\text{Var}^\pi\left((r_i(c))_{c \in \mathcal{C}}\right)\right)\right),$$

$$\text{Var}^\pi\left((r_i(c))_{c \in \mathcal{C}}\right) = \sum_{c \in \mathcal{C}} \pi(c)\left(r_i(c) - \sum_{c_1 \in \mathcal{C}} \pi(c_1)r_i(c_1)\right)^2.$$

The optimization problem OPTSTAT given below:

$$\max_{(\mathbf{r}(c))_{c \in \mathcal{C}}} \quad \phi_\pi \left( (\mathbf{r}(c))_{c \in \mathcal{C}} \right),$$

$$\text{subject to} \quad c(\mathbf{r}(c)) \leq 1, \quad \forall \ c \in \mathcal{C},$$

$$r_i(c) \geq r_{\min}, \quad \forall \ c \in \mathcal{C}.$$

The next result gives few useful properties of OPTSTAT.

**Lemma 2.** *(a) OPTSTAT is a convex optimization problem satisfying Slater's condition.*
*(b) OPTSTAT has a unique solution.*

*Proof:* The proof is similar to that of Lemma 1 (and is easy to establish once we prove the convexity of the function $\text{Var}^\pi(.)$).  ∎

Using Lemma 2 (a), we can conclude that KKT conditions are necessary and sufficient for optimality for OPTSTAT. Let $(\mathbf{r}^\pi(c) : c \in \mathcal{C})$ denote the optimal solution.

**KKT-OPTSTAT**:
There exist constants $(\mu^\pi(c) : c \in \mathcal{C})$ and $\left( (\gamma_i^\pi(c))_{i \in \mathcal{N}} : c \in \mathcal{C} \right)$ are such that

$$\pi(c) \left( \left( U_i^R \right)' (r_i^\pi(c)) \right.$$

$$\left. - 2 \left( U_i^V \right)' \left( \text{Var}^\pi \left( (r_i^\pi(c))_{c \in \mathcal{C}} \right) \right) \left( r_i^\pi(c) - \sum_{c \in \mathcal{C}} \pi(c) r_i^\pi(c) \right) \right)$$

$$- \mu^\pi(c) c_i'(\mathbf{r}^\pi(c)) + \gamma_i^\pi(c) = 0, \quad (9)$$

$$\mu^\pi(c) c(\mathbf{r}^\pi(c)) = 0, \quad (10)$$

$$\gamma_i^\pi(c) (r_i^\pi(c) - r_{\min}) = 0, \quad (11)$$

where $c_i'$ denotes $\frac{\partial c}{\partial r_i}$, and we used following result: for any $c_0 \in \mathcal{C}, \ i \in \mathcal{N}$,

$$\frac{\partial \text{Var}^\pi \left( (r_i(c))_{c \in \mathcal{C}} \right)}{\partial r_i(c_0)} = 2\pi(c_0) \left( r_i(c_0) - \sum_{c_1 \in \mathcal{C}} \pi(c_1) r_i(c_1) \right).$$

## V. Adaptive Variance aware Reward allocation

In this section, we present our online algorithm AVR to solve $\text{OPT}(T)$, and establish its asymptotic optimality.

The reward allocations for AVR are obtained by solving OPTAVR$(\mathbf{m}, \mathbf{v}, c)$ given below:

$$\max_{\mathbf{r}} \sum_{i \in \mathcal{N}} \left( U_i^R(r_i) - \left( U_i^V \right)' (v_i)(r_i - m_i)^2 \right) + h_0(\mathbf{v})$$

$$\text{subject to} \quad c(\mathbf{r}) \leq 0, \quad (12)$$

$$r_i \geq r_{\min} \ \forall \ i \in \mathcal{N}, \quad (13)$$

where

$$h_0(\mathbf{v}) = \sum_{i \in \mathcal{N}} \left( \left( U_i^V \right)' (v_i) v_i - U_i^V(v_i) \right).$$

Note that OPTAVR$(\mathbf{m}, \mathbf{v}, c)$ is closely related to OPT-ONLINE (discussed in Subsection I-A). Also, note that $h_0(\mathbf{e}, \mathbf{v})$ does not depend on the allocation and thus can be ignored while solving the optimization problem. But, it modifies the objective function and (thus) the optimal value of the objective function to ensure certain nice properties for the partial derivatives of latter (see Lemma 3 (b)). Let $\mathbf{r}^*(\mathbf{m}, \mathbf{v}, c)$ denote the optimal solution to OPTAVR$(\mathbf{m}, \mathbf{v}, c)$. Also, let $\mathcal{H}$ be given by:

$$\mathcal{H} = [r_{\min}, r_{\max}]^N \times [0, v_{\max}]^N,$$

where $\times$ denotes cross product operator for sets.

Next, we describe the algorithm AVR in detail. AVR consists of three steps, AVR.0-AVR.2, given next:

---

**Adaptive Variance aware Reward allocation (AVR)**

---

**AVR.0**: Initialize: Let $(\widehat{\mathbf{m}}(0), \widehat{\mathbf{v}}(0)) \in \mathcal{H}$.

In each slot $t + 1$ for $t \geq 0$, carry out the following steps:
**AVR.1**: The reward allocation in slot $t$ is given by $\mathbf{r}^*(\widehat{\mathbf{m}}(t), \widehat{\mathbf{e}}(t), \widehat{\mathbf{v}}(t), c_{t+1})$ and will be denoted by $\mathbf{r}^*(t+1)$ (when the dependence on the variables is clear from context).
**AVR.2**: In slot $t$, update $\widehat{m}_i$ as follows: for all $i \in \mathcal{N}$,

$$\widehat{m}_i(t+1) = \widehat{m}_i(t) + \frac{1}{t} \left( r_i^*(t+1) - \widehat{m}_i(t) \right), \quad (14)$$

and update $\widehat{v}_i$ as follows: for all $i \in \mathcal{N}_{Vl}$, $\widehat{v}_i(t+1) = \widehat{v}_i(0)$, and for all $i \in \mathcal{N}_{Vn}$,

$$\widehat{v}_i(t+1) = \widehat{v}_i(t) + \frac{1}{t} \left( (r_i^*(t+1) - \widehat{m}_i(t))^2 - \widehat{v}_i(t) \right). \quad (15)$$

---

We see that the update equations (14)-(15) roughly ensure that the parameters $\widehat{\mathbf{m}}(t)$ and $(\widehat{v}_i(t))_{i \in \mathcal{N}_{Vn}}$ keep track of mean reward and variance in reward respectively associated with the reward allocation under AVR. Also, note that we do not have to keep track of the estimates of variance in reward seen by users $i$ with linear $U_i^V$.

We let $\widehat{\boldsymbol{\theta}}_t = (\widehat{\mathbf{m}}(t), \widehat{\mathbf{v}}(t))$ for each $t$. The update equations (14)-(15) ensure that $\widehat{\boldsymbol{\theta}}_t$ stays in the set $\mathcal{H}$.

For any $(\mathbf{m}, \mathbf{v}, c) \in \mathcal{H}$, we have $\left( U_i^V \right)' (v_i) > 0$ (see assumption U.V). Hence, OPTAVR$(\mathbf{m}, \mathbf{v}, c)$ is a convex optimization problem with a unique solution. Further, using assumption C.4, we can show that it satisfies Slater's condition. Hence, the optimal solution for OPTAVR$(\mathbf{m}, \mathbf{v}, c)$ satisfies KKT conditions given below.

**KKT-OPTAVR$(\mathbf{m}, \mathbf{v}, c)$**:
There exist non-negative constants $\mu^*$ and $(\gamma_i^* : i \in \mathcal{N})$ such that for all $i \in \mathcal{N}$

$$\left( U_i^R \right)' (r_i^*) - 2 \left( U_i^V \right)' (v_i)(r_i^* - m_i)$$

$$+ \gamma_i^* - \mu^* c_i'(\mathbf{r}^*) = 0, \quad (16)$$

$$\mu^* c(\mathbf{r}^*) = 0, \quad (17)$$

$$\gamma_i^* (r_i^* - r_{\min}) = 0. \quad (18)$$

---

Let $h(\mathbf{m}, \mathbf{v}, c)$ denote the optimal value of the objective function of OPTAVR$(\mathbf{m}, \mathbf{v}, c)$, i.e., $h$ is a function defined on an open interval (the obvious one that can be obtained from the domains of the functions $\left(U_i^R, U_i^V\right)_{i \in \mathcal{N}}$ containing $\mathcal{H}$ as given below

$$
\begin{aligned}
h\left(\mathbf{m}, \mathbf{v}, c\right) = & \sum_{i \in \mathcal{N}}\Big(U_i^R\left(r_i^*\right) - U_i^V\left(v_i\right) \\
& + \left(U_i^V\right)'\left(v_i\right)\left(v_i - \left(r_i^* - m_i\right)^2\right)\Big),
\end{aligned}
$$

where $\mathbf{r}^*$ stands for $\mathbf{r}^*\left(\mathbf{m}, \mathbf{v}, c\right)$.

In the next result, we establish continuity and differentiability properties of $\mathbf{r}^*\left(\mathbf{m}, \mathbf{v}, c\right)$ (also denoted by $\mathbf{r}^*$ in the result) and $h\left(\mathbf{m}, \mathbf{v}, c\right)$ respectively, viewing them as functions of $\left(\mathbf{m}, \mathbf{v}\right)$.

**Lemma 3.** *For any $c \in \mathcal{C}$, and $\boldsymbol{\theta} = \left(\mathbf{m}, \mathbf{v}\right) \in \mathcal{H}$*
*(a) $\mathbf{r}^*\left(\boldsymbol{\theta}, c\right)$ is a continuous function of $\boldsymbol{\theta}$.*
*(b) For each $i \in \mathcal{N}$,*

$$
\begin{aligned}
\frac{\partial h\left(\boldsymbol{\theta}, c\right)}{\partial m_i} &= 2\left(r_i^* - m_i\right)\left(U_i^V\right)'\left(v_i\right), \\
\frac{\partial h\left(\boldsymbol{\theta}, c\right)}{\partial v_i} &= \left(U_i^V\right)''\left(v_i\right)\left(v_i - \left(r_i^* - m_i\right)^2\right).
\end{aligned}
$$

*(c) $E\left[\mathbf{r}^*\left(\boldsymbol{\theta}, C^\pi\right)\right]$ is a continuous function of $\boldsymbol{\theta}$.*
*(d) For each $i \in \mathcal{N}$,*

$$
\begin{aligned}
\frac{\partial E\left[h\left(\boldsymbol{\theta}, C^\pi\right)\right]}{\partial m_i} &= 2\left(E\left[r_i^*\left(\boldsymbol{\theta}, C^\pi\right)\right] - m_i\right)\left(U_i^V\right)'\left(v_i\right), \\
\frac{\partial E\left[h\left(\boldsymbol{\theta}, C^\pi\right)\right]}{\partial v_i} &= \left(U_i^V\right)''\left(v_i\right) \\
& \quad \left(v_i - E\left[\left(r_i^*\left(\boldsymbol{\theta}, C^\pi\right) - m_i\right)^2\right]\right).
\end{aligned}
$$

Proof Sketch: Proofs of parts (a) and (b) mainly rely on some fundamental results on perturbation analysis of optimization problems from [?] and [?]. Part (a) can be proved using Theorem 2.2 in [?]. The result in part (b) can be shown using Theorem 4.1 in [?]. This theorem tells us that if certain conditions are met, then we can evaluate the partial derivative of the optimal value of a parametric optimization problem (with respect to any parameter) by just evaluating the partial derivative of the objective of the optimization problem, and then substituting the optimal solution. For instance, by using the theorem, we can evaluate the partial derivative of the optimal value $h\left(\boldsymbol{\theta}, c\right)$ with respect to $m_i$ as follows. We first evaluate the partial derivative of the objective function of OPTAVR $\left(\boldsymbol{\theta}, c\right)$:

$$
\begin{aligned}
\frac{\partial}{\partial m_i}\Bigg( & h_0\left(\mathbf{v}\right) + \sum_{i \in \mathcal{N}}\left(U_i^R\left(r_i\right) - \left(U_i^V\right)'\left(v_i\right)\left(r_i - m_i\right)^2\right)\Bigg) \\
& = 2\left(r_i - m_i\right)\left(U_i^V\right)'\left(v_i\right).
\end{aligned}
$$

Now, on substituting $\mathbf{r}^*$ in the above expression, we obtain the first result in part (b). The other results can be obtained similarly.

Parts (c) and (d) can shown using parts (a) and (b) respectively, and Bounded Convergence Theorem (see [?]). $\quad\square$

From part (b) of the above result, we see that the update equations (14)-(15) ensure that $\widehat{\boldsymbol{\theta}}(t)$ moves in a direction that increases $h(.)$. This is in part due to the careful choice of the function $h_0$ (which is independent of variables being optimized) appearing in the objective function of OPTAVR.

Next, we find relationships between the optimal solution $\left(\mathbf{r}^\pi\left(c\right) : c \in \mathcal{C}\right)$ of OPTSTAT and OPTAVR. Towards that end, let $m_i^\pi = \sum_{c \in \mathcal{C}} \pi\left(c\right) r_i^\pi\left(c\right)$ and $v_i^\pi = \text{Var}^\pi\left(\left(r_i^\pi\left(c\right)\right)_{c \in \mathcal{C}}\right)$ for each $i \in \mathcal{N}$. Next, let

$$
\mathcal{H}^* = \left\{\left(\mathbf{m}, \mathbf{v}\right) \in \mathcal{H} : \left(\mathbf{m}, \mathbf{v}\right) \text{ satisfies } (19) - (20)\right\},
$$

where the conditions (19)-(20) are given below:

$$
\begin{aligned}
E\left[r_i^*\left(\mathbf{m}, \mathbf{v}, C^\pi\right)\right] &= m_i & \forall \ i \in \mathcal{N}, & \quad (19) \\
\text{Var}\left(r_i^*\left(\mathbf{m}, \mathbf{v}, C^\pi\right)\right) &= v_i & \forall \ i \in \mathcal{N}_{Vn}. & \quad (20)
\end{aligned}
$$

Part (a) of the next result provides a fixed point like relationship for the optimal solution to OPTSTAT using the optimal solution function $\mathbf{r}^*(.)$ of OPTAVR, and part (b) is a useful consequence of part (a). A proof for the result is given in Appendix A.

**Lemma 4.** $\left(\mathbf{m}^\pi, \mathbf{v}^\pi\right)$ *satisfies*
*(a) $\mathbf{r}^*\left(\mathbf{m}^\pi, \mathbf{v}^\pi, c\right) = \mathbf{r}^\pi\left(c\right)$ for each $c \in \mathcal{C}$, and*
*(b) $\left(\mathbf{m}^\pi, \mathbf{v}^\pi\right) \in \mathcal{H}^*$.*

The next result tells us that we can obtain the optimal solution to OPTSTAT from any element in $\mathcal{H}^*$ by using the optimal solution function $\mathbf{r}^*(.)$. Further, it gives us very useful uniqueness results for the components of the elements in $\mathcal{H}^*$. A proof for the result is given in Appendix B.

**Lemma 5.** *Suppose $\left(\mathbf{m}_1, \mathbf{v}_1\right) \in \mathcal{H}^*$. Then,*
*(a) $\left(\mathbf{r}^*\left(\mathbf{m}_1, \mathbf{v}_1, c\right)\right)_{c \in \mathcal{C}}$ is an optimal solution to OPTSTAT. Suppose that $\left(\mathbf{m}_2, \mathbf{v}_2\right) \in \mathcal{H}^*$. Then,*
*(b) $\mathbf{r}^*\left(\mathbf{m}_1, \mathbf{v}_1, c\right) = \mathbf{r}^*\left(\mathbf{m}_2, \mathbf{v}_2, c\right)$ for each $c \in \mathcal{C}$, and*
*(c) $m_{1i} = m_{2i}$ for each $i \in \mathcal{N}$, and $v_{1i} = v_{2i}$ for each $i \in \mathcal{N}_{Vn}$.*
*(d) $m_{1i} = m_i^\pi$ for each $i \in \mathcal{N}$, and $v_{1i} = v_i^\pi$ for each $i \in \mathcal{N}_{Vn}$.*

Till now, we focused only on the optimization problem OPTAVR associated with AVR. In the next subsection, we study the evolution of $\left(\widehat{\boldsymbol{\theta}}_t\right)_t$ under AVR.

### A. Convergence Analysis

In this subsection, we focus on establishing some properties related to the convergence of the sequence $\left(\widehat{\boldsymbol{\theta}}_t\right)_t$ that are key to proof of the main optimality result (Theorem 1).

Towards that end, we study the the differential equation

$$
\frac{d\boldsymbol{\theta}(t)}{dt} = \bar{\mathbf{g}}\left(\boldsymbol{\theta}(t)\right), \quad (21)
$$

where $\bar{\mathbf{g}}\left(\boldsymbol{\theta}\right)$ is a function taking values in $\mathbb{R}^{3N}$ defined as follows: for $\boldsymbol{\theta} = \left(\mathbf{m}, \mathbf{v}\right) \in \mathcal{H}$, let

$$
\begin{aligned}
\left(\bar{\mathbf{g}}\left(\boldsymbol{\theta}\right)\right)_i &= E\left[r_i^*\left(\boldsymbol{\theta}, C^\pi\right)\right] - m_i, \\
\left(\bar{\mathbf{g}}\left(\boldsymbol{\theta}\right)\right)_{N+i} &= \mathbf{I}_{\{i \in \mathcal{N}_{Vn}\}}\left(E\left[\left(r_i^*\left(\boldsymbol{\theta}, C^\pi\right) - m_i\right)^2\right] - v_i\right).
\end{aligned}
$$

The motivation for studying the above differential equation should be partly clear by comparing the RHS of (21) with the update equations in (14)-(15) in AVR.

Now we study (21) in light of the above result and obtain a convergence result for the differential equation, which tells us that for any initial condition, $\boldsymbol{\theta}(t)$ evolving according to (21) converges to the set $\mathcal{H}_*$ given by

$$\mathcal{H}_* = \{ \boldsymbol{\theta} = (\mathbf{m}^\pi, \mathbf{v}) : \mathbf{r}^*(\boldsymbol{\theta}, c) = \mathbf{r}^*(\boldsymbol{\theta}^\pi, c) \ \forall \ c \in \mathcal{C},$$
$$\left(U_i^V\right)'(v_i) = \left(U_i^V\right)'(v_i^\pi) \ \forall \ i \in \mathcal{N}_{Vn} \}$$

We can verify that $\mathcal{H}^* \subset \mathcal{H}_*$ (using (19)-(20)). A proof for the next result is discussed in Appendix C.

**Lemma 6.** *Suppose $\boldsymbol{\theta}(t)$ evolves according to (21). Then, $\boldsymbol{\theta}(t)$ converges to $\mathcal{H}_*$ as $t$ tends to infinity for any $\boldsymbol{\theta}(0) \in \mathcal{H}$.*

Now, due to the above result, we have a key convergence result for the differential equation (21) which is closely related to the update equations (14)-(15) of AVR. Next, we use this result to obtain a convergence result for $\left(\widehat{\boldsymbol{\theta}}_t\right)_t$. We do so by viewing (14)-(15) as a stochastic approximation update equation, and using a result from [?] that helps us to relate it the differntial equation (21).

**Lemma 7.** *If $\widehat{\boldsymbol{\theta}}_0 \in \mathcal{H}$, then the sequence $\left(\widehat{\boldsymbol{\theta}}_t\right)_t$ generated by AVR converges almost surely to the set $\mathcal{H}_*$.*

Proof Sketch: We can prove the result by viewing (14)-(15) as a stochastic approximation update equation. The proof mainly uses Lemma 6 and Theorem 1.1 of Chapter 6 from [?] (that gives sufficient conditions for convergence of a stochastic approximation scheme). □

We had pointed out that our main interest is in the convergence properties of $\left(\widehat{m}_i(t), \left(U_i^V\right)'(\widehat{v}_i(t))\right)_{i \in \mathcal{N}}$. The next result uses Lemma 7 to establish the desired convergence property. A proof for the result is given in Appendix D.

**Lemma 8.** *If $\widehat{\boldsymbol{\theta}}_0 \in \mathcal{H}$, then the sequence $\left(\widehat{\boldsymbol{\theta}}_t\right)_t$ generated by AVR satisfies:*
*(a) For each $i \in \mathcal{N}$, $\lim_{t \to \infty} \widehat{\mathbf{m}}(t) = \mathbf{m}^\pi$, and*
*(b) $\lim_{t \to \infty} \mathbf{r}^*\left(\widehat{\boldsymbol{\theta}}(t), c\right) = \mathbf{r}^*(\boldsymbol{\theta}^\pi, c)$, and*
*(c) For each $i \in \mathcal{N}_{Vn}$, $\lim_{t \to \infty} \left(U_i^V\right)'(\widehat{v}_i(t)) = \left(U_i^V\right)'(v_i^\pi)$.*

Next, we use Lemma 8 and stationarity to establish certain properties associated with the time averages of the reward allocations under the online scheme AVR. For brevity, in the following result, we let $\mathbf{r}^*(t)$ denote $\mathbf{r}^*(\widehat{\mathbf{m}}(t), \widehat{\mathbf{v}}(t), c_t)$. A proof for the result is given in Appendix E.

**Lemma 9.** *For almost all sample paths,*

(a) *For each $i \in \mathcal{N}$, $\displaystyle\lim_{T \to \infty} \frac{1}{T} \sum_{\tau=1}^T r_i^*(\tau) = \lim_{t \to \infty} \widehat{m}_i(t)$.*

(b) *For each $i \in \mathcal{N}$,*
$$\lim_{T \to \infty} \left(U_i^V\right)' \left(Var^T\left((r_i^*)_{1:T}\right)\right) = \lim_{t \to \infty} \left(U_i^V\right)'(\widehat{v}_i(t)).$$

### B. Asymptotic Optimality of AVR

The next result establishes the asymptotic optimality of AVR, i.e., if we run AVR for long enough period, the difference in performance of AVR and the optimal finite horizon policy becomes negligible.

**Theorem 1.** *For almost all sample paths the following two statements hold:*
*(a) Feasibility: The allocation $(\mathbf{r}^*)_{1:T}$ associated with AVR satisfies (4) and (5), and for each $i \in \mathcal{N}$.*
*(b) Optimality: AVR is asymptotically optimal, i.e.,*

$$\lim_{T \to \infty} \left(\phi_T\left((\mathbf{r}^*)_{1:T}\right) - \phi_T\left((\mathbf{r^T})_{1:T}\right)\right) = 0.$$

Proof: Since the allocation $(\mathbf{r}^*)_{1:T}$ associated with AVR satisfies (12) and (13) in each time slot, it also satisfies (4) and (5). Thus, part (a) is true.

To prove part (b), consider any realization of $(c)_{1:T}$. Let $(\mu^*)_{1:T}$ and $(\gamma_i^* : i \in \mathcal{N})_{1:T}$ be the sequences of non negative real numbers satisfying (16), (17) and (18) for the realization. Hence, from the non-negativity of these numbers, and feasibility of $(\mathbf{r}^T)_{1:T}$, we have

$$\phi_T\left((\mathbf{r}^T)_{1:T}\right) \ \leq \ \varphi_T\left((\mathbf{r}^T)_{1:T}\right).$$

where

$$\begin{aligned}
\varphi_T\left((\mathbf{r}^T)_{1:T}\right) \ = \ & \sum_{i \in \mathcal{N}} \left(\frac{1}{T} \sum_{t=1}^T U_i^R\left(r_i^T(t)\right)\right. \\
& \left. - U_i^V\left(\mathrm{Var}^T\left((r_i^T)_{1:T}\right)\right)\right) \\
& - \sum_{t=1}^T \frac{\mu^*(t)}{T} c_t(\mathbf{r}^T(t)) \\
& + \sum_{t=1}^T \sum_{i \in \mathcal{N}} \frac{\gamma_i^*(t)}{T} \left(r_i^T(t) - r_{\min}\right).
\end{aligned}$$

Since $\varphi_T$ is a differentiable concave function, we have (see [?])

$$\begin{aligned}
\varphi_T\left((\mathbf{r}^T)_{1:T}\right) \ \leq \ & \varphi_T((\mathbf{r}^*)_{1:T}) \\
& + \nabla\varphi_T((\mathbf{r}^*)_{1:T}) \bullet \left((\mathbf{r}^T)_{1:T} - (\mathbf{r}^*)_{1:T}\right),
\end{aligned}$$

where '•' denotes the dot product. Hence, we have

$$\phi_T\left((\mathbf{r}^T)_{1:T}\right) \le \varphi_T\left((\mathbf{r}^T)_{1:T}\right)$$

$$\le \sum_{i \in \mathcal{N}} \left( \frac{1}{T} \sum_{t=1}^{T} U_i^R\left(r_i^*(t)\right) - U_i^V\left(\mathrm{Var}^T\left((r_i^*)_{1:T}\right)\right) \right)$$

$$- \sum_{t=1}^{T} \frac{\mu^*(t)}{T} c_t(\mathbf{r}^*(t))$$

$$+ \sum_{t=1}^{T} \sum_{i \in \mathcal{N}} \frac{\gamma_i^*(t)}{T} \left(r_i^*(t) - r_{\min}\right)$$

$$+ \sum_{t=1}^{T} \sum_{i \in \mathcal{N}} \left(r_i^T(t) - r_i^*(t)\right)$$

$$\left( \frac{\left(U_i^R\right)'\left(r_i^*(t)\right)}{T} \right.$$

$$- \frac{2\left(U_i^V\right)'\left(\mathrm{Var}^T\left((r_i^*)_{1:T}\right)\right)}{T} \left(r_i^*(t) - \frac{1}{T}\sum_{\tau=1}^{T} r_i^*(\tau)\right)$$

$$\left. - \frac{\mu^*(t)}{T} c_{t,i}'(\mathbf{r}^*(t)) + \frac{\gamma_i^*(t)}{T}\right).$$

Now, since $(\mu^*)_{1:T}$ and $(\gamma_i^* : i \in \mathcal{N})_{1:T}$ satisfy (16), (17) and (18), we have

$$\phi_T\left((\mathbf{r}^T)_{1:T}\right) \tag{22}$$

$$\le \sum_{i \in \mathcal{N}} \left( \frac{1}{T} \sum_{t=1}^{T} U_i^R\left(r_i^*(t)\right) - U_i^V\left(\mathrm{Var}^T\left((r_i^*)_{1:T}\right)\right) \right)$$

$$+ \sum_{t=1}^{T} \sum_{i \in \mathcal{N}} \frac{r_i^T(t) - r_i^*(t)}{T}$$

$$\left( \left(U_i^R\right)'\left(r_i^*(t)\right) - \left(U_i^R\right)'\left(r_i^*(t)\right) \right.$$

$$- 2\left(U_i^V\right)'\left(\mathrm{Var}^T\left((r_i^*)_{1:T}\right)\right) \left(r_i^*(t) - \frac{1}{T}\sum_{\tau=1}^{T} r_i^*(\tau)\right)$$

$$\left. + 2\left(U_i^V\right)'\left(\widehat{v}_i(t-1)\right)\left(r_i^*(t) - \widehat{m}_i(t-1)\right)\right).$$

From Lemma 9 (a)-(c) and the continuity of the functions involved, we can conclude that the following term (appearing above) can be made as small as desired by choosing large enough $T$ and then choosing a large enough $t$:

$$\left( \left(U_i^R\right)'\left(r_i^*(t)\right) - \left(U_i^R\right)'\left(r_i^*(t)\right) \right.$$

$$- 2\left(U_i^V\right)'\left(\mathrm{Var}^T\left((r_i^*)_{1:T}\right)\right) \left(r_i^*(t) - \frac{1}{T}\sum_{\tau=1}^{T} r_i^*(\tau)\right)$$

$$\left. + 2\left(U_i^V\right)'\left(\widehat{v}_i(t-1)\right)\left(r_i^*(t) - \widehat{m}_i(t-1)\right)\right).$$

Also, $\left|r_i^T(t) - r_i^*(t)\right| \le r_{\max}$. Hence, taking limits in (22),

$$\lim_{T \to \infty} \frac{1}{T}\left(\phi_T\left((\mathbf{r}^*)_{1:T}\right) - \phi_T\left((\mathbf{r}^T)_{1:T}\right)\right) \ge 0.$$

holds for almost all sample paths. From optimality of $(\mathbf{r}^T)_{1:T}$,

$$\phi_T\left((\mathbf{r}^T)_{1:T}\right) \ge \phi_T\left((\mathbf{r}^*)_{1:T}\right).$$

From the above two inequalities, the result follows. ■

## VI. Conclusions

The two main contributions of this work are summarized below:
(1) We propose a novel framework for reward allocation to users who are sensitive to temporal variability in the reward allocation. The formulation allows tradeoffs between mean and variability associated with the reward allocation of the users by appropriately choosing the functions $\left(U_i^R, U_i^V\right)_{i \in \mathcal{N}}$.
(2) We proposed an asymptotically optimal online algorithm AVR to solve problems falling in this framework.

## Appendix A
### Proof of Lemma 4

For each $c \in \mathcal{C}$, by choosing $\mathbf{r}^*\left(\mathbf{m}^\pi, \mathbf{v}^\pi, c\right) = \mathbf{r}^\pi(c)$, $\mu^* = \frac{\mu^\pi(c)}{\pi(c)}$ and $\gamma_i^* = \frac{\gamma_i^\pi(c)}{\pi(c)}$ for all $i \in \mathcal{N}$, we can verify that $\mathbf{r}^*\left(\mathbf{m}^\pi, \mathbf{e}^\pi, \mathbf{v}^\pi, c\right)$ along with $\mu^*$ and $(\gamma_i^* : i \in \mathcal{N})$ satisfy (16)-(18) using the fact that $(\mathbf{r}^\pi(c) : c \in \mathcal{C})$, $(\mu^\pi(c) : c \in \mathcal{C})$ and $\left((\gamma_i^\pi(c))_{i \in \mathcal{N}} : c \in \mathcal{C}\right)$ satisfy (9)-(11).

Part (b) follows from the definitions of $\mathbf{m}^\pi$ and $\mathbf{v}^\pi$.

## Appendix B
### Proof of Lemma 5

For each $c \in \mathcal{C}$, $\mathbf{r}^*\left(\mathbf{m}_1, \mathbf{v}_1, c\right)$ is an optimal solution to OPTAVR and thus, there exist (like those in KKT-OPTAVR given in (16)-(18)) non-negative constants $\mu_1^*(c)$ and $(\gamma_{1i}^*(c) : i \in \mathcal{N})$ such that for all $i \in \mathcal{N}$,

$$\left(U_i^R\right)'\left(r_i^*(c)\right) - \left(U_i^V\right)'(v_{1i})\left(r_i^*(c) - m_{1i}^*\right)$$
$$+ \gamma_{1i}^* - \mu_1^*(c)\, c_i'(\mathbf{r}^*(c)) = 0,$$
$$\mu_1^*(c)\, c(\mathbf{r}^*(c)) = 0,$$
$$\gamma_{1i}^*\left(r_i^*(c) - r_{\min}\right) = 0,$$

where we used $\mathbf{r}^*(c)$ instead of $\mathbf{r}^*\left(\mathbf{m}_1, \mathbf{v}_1, c\right)$ for brevity.

For each $i \in \mathcal{N}_{VI}$, due to linearity we have that $\left(U_i^V\right)'(.)$ is a constant, and hence is independent of its argument. Thus, we have $\left(U_i^V\right)'(v_{1i}) = \left(U_i^V\right)'\left(\mathrm{Var}\left(\mathbf{r}^*(C^\pi)\right)\right)$. Further, note that $\left(\mathbf{m}_1, \mathbf{v}_1\right) \in \mathcal{H}^*$ and hence satisfies (19)-(20). Using these arguments, we can rewrite the above equations as follows: for all $c \in \mathcal{C}$

$$\left(U_i^R\right)'\left(r_i^*(c)\right) - \left(U_i^V\right)'\left(\mathrm{Var}^\pi\left(\mathbf{r}^*(C^\pi)\right)\right)\left(r_i^*(c) \right.$$
$$\left. - E\left[\mathbf{r}^*(C^\pi)\right]\right) + \gamma_{1i}^* - \mu_1^*(c)\, c_i'(\mathbf{r}^*(c)) = 0,$$
$$\mu_1^*(c)\, c(\mathbf{r}^*(c)) = 0,$$
$$\gamma_{1i}^*\left(r_i^*(c) - r_{\min}\right) = 0,$$

Now for each $c \in \mathcal{C}$, multiply the above equations with $\pi(c)$ and one obtains KKT-OPTSTAT ((9)-(11)) with $(\pi(c)\mu_1^*(c) : c \in \mathcal{C})$ and $\left((\pi(c)\gamma_{1i}^*(c))_{i \in \mathcal{N}} : c \in \mathcal{C}\right)$ as Lagrange multipliers. From Lemma 2 (a), OPTSTAT satisfies Slater's condition and hence KKT conditions are sufficient for optimality of OPTSTAT. Thus, we have that $\left(\mathbf{r}^*\left(\mathbf{m}_1, \mathbf{v}_1, c\right)\right)_{c \in \mathcal{C}}$ is an optimal solution to OPTSTAT. This proves part (a).

Now suppose that $\left(\mathbf{m}_1, \mathbf{v}_1\right), \left(\mathbf{m}_2, \mathbf{v}_2\right) \in \mathcal{H}^*$, and suppose that for some $c_0 \in \mathcal{C}$ and $i \in \mathcal{N}$, $r_i^*\left(\mathbf{m}_1, \mathbf{v}_1, c_0\right) \ne r_i^*\left(\mathbf{m}_2, \mathbf{v}_2, c_0\right)$. Thus, using this together with part (a), we have that $(\mathbf{r}^*(\mathbf{m}_1, \mathbf{v}_1, c))_{c \in \mathcal{C}}$ and $(\mathbf{r}^*(\mathbf{m}_2, \mathbf{v}_2, c))_{c \in \mathcal{C}}$ are two distinct solutions to OPTSTAT. However, this contradicts fact that OPTSTAT has a unique solution (see Lemma 2(b)). Thus, (b) has to hold.

Now suppose that $(\mathbf{m}_1, \mathbf{v}_1), (\mathbf{m}_2, \mathbf{v}_2) \in \mathcal{H}^*$. and that (c) does not hold. Then, we can conclude that atleast one of the conditions given in part (c) does not hold. For instance, suppose that $v_{1j} \neq v_{2j}$ for some $j \in \mathcal{N}_{Vn}$. This along with the fact that $(\mathbf{m}_1, \mathbf{v}_1), (\mathbf{m}_2, \mathbf{v}_2) \in \mathcal{H}^*$ (and thus they satisfy (20)) implies that $\text{Var}(r_i^*(\mathbf{m}_1, \mathbf{v}_1, C^\pi)) \neq \text{Var}(r_i^*(\mathbf{m}_2, \mathbf{v}_2, C^\pi))$. Thus, we can conclude that for some $c_0 \in \mathcal{C}$ and $i \in \mathcal{N}$, $r_i^*(\mathbf{m}_1, \mathbf{v}_1, c_0) \neq r_i^*(\mathbf{m}_2, \mathbf{v}_2, c_0)$. We can reach the same conclusion if any of the conditions given in (c) are violated. But, the conclusion contradicts part (b). Thus, (c) has to hold.

Part (d) follows from part (c) and Lemma 4 part (b).

## Appendix C
## Proving Lemma 6

We let $\boldsymbol{\theta}^\pi = (\mathbf{m}^\pi, \mathbf{e}^\pi, \mathbf{v}^\pi)$, and $\boldsymbol{\theta} = (\mathbf{m}, \mathbf{v})$, and consider the Lyapunov function $V(\boldsymbol{\theta}) = E[h(\boldsymbol{\theta}^\pi, C^\pi)] - E[h(\boldsymbol{\theta}, C^\pi)]$.
Then

$$
\begin{aligned}
\frac{dV(\boldsymbol{\theta}(t))}{dt} &= \nabla V(\boldsymbol{\theta}(t)) \cdot \frac{d\boldsymbol{\theta}(t)}{dt} \\
&= \nabla V(\boldsymbol{\theta}(t)) \cdot (\bar{\mathbf{g}}(\boldsymbol{\theta}(t)) + \mathbf{z}(\boldsymbol{\theta}(t))) \\
&= \nabla V(\boldsymbol{\theta}(t)) \cdot \bar{\mathbf{g}}(\boldsymbol{\theta}(t)),
\end{aligned}
$$

where the last step follows from Lemma ??. Let $\dot{V}(\boldsymbol{\theta}) = \nabla V(\boldsymbol{\theta}) \cdot \bar{\mathbf{g}}(\boldsymbol{\theta})$. Then from Lemma 3 (d) and Lemma ??, we have that for any $\boldsymbol{\theta} \in \mathcal{H}$,

$$
\begin{aligned}
\dot{V}(\boldsymbol{\theta}) &= -\sum_{i \in \mathcal{N}} 2\left(U_i^V\right)'(v_i)\left(E[r_i^*(\boldsymbol{\theta}, C^\pi)] - m_i\right)^2 \\
&+ \sum_{i \in \mathcal{N}_{Vn}} \left(U_i^V\right)''(v_i)\left(v_i - E\left[(r_i^*(\boldsymbol{\theta}, C^\pi) - m_i)^2\right]\right)^2.
\end{aligned}
$$

The expression above is the negative of a sum of (positive) weighted squares. Hence,

$$
\dot{V}(\boldsymbol{\theta}) \leq 0 \quad \forall \ \boldsymbol{\theta} \in \mathcal{H}. \tag{23}
$$

Now, let $\mathcal{H}_V = \left\{\boldsymbol{\theta} \in \mathcal{H} : \dot{V}(\boldsymbol{\theta}) = 0\right\}$. Since $V(.)$ is a continuously differentiable function on the (compact) set $\mathcal{H}$ satisfying (23), we can use LaSalle's Theorem (see Theorem 4.4 in [?]) to conclude that $\boldsymbol{\theta}(t)$ converges to the largest invariant set in $\mathcal{H}_V$. Let $\mathcal{H}_\pi$ denote the set.

In the remaining part of the proof, we prove that $\mathcal{H}_\pi \subset \mathcal{H}_*$ from which the main claim follows.

Noting that $\dot{V}(\boldsymbol{\theta}) = 0$ for any $\boldsymbol{\theta} \in \mathcal{H}_\pi$, and using the expression for $\dot{V}()$ given above, we can show that

$$
E[\mathbf{r}^*(\boldsymbol{\theta}, C^\pi)] = \mathbf{m} \quad \forall \ \boldsymbol{\theta} \in \mathcal{H}_\pi. \tag{24}
$$

Also, for any $\boldsymbol{\theta} \in \mathcal{H}_\pi$, $\dot{V}(\boldsymbol{\theta}) = 0$, and hence using the fact that $\min_{v \in [0, v_{\max}]} \left(U_i^V\right)''(v) < 0$, we have that

$$
v_i = E\left[(r_i^*(\boldsymbol{\theta}, C^\pi) - m_i)^2\right] \quad \forall \ i \in \mathcal{N}_{Vn}.
$$

From the above conclusion and (24), we can conclude that for any $\boldsymbol{\theta} \in \mathcal{H}_\pi$, we have $\boldsymbol{\theta} \in \mathcal{H}^*$. Since $\mathcal{H}^* \subset \mathcal{H}_*$, we have that for $\mathcal{H}_\pi \subset \mathcal{H}_*$. Now, since $\boldsymbol{\theta}(t)$ converges to $\mathcal{H}_\pi$, we can conclude that $\boldsymbol{\theta}(t)$ converges to $\mathcal{H}_*$ and the result follows.

## Appendix D
## Proof for Lemma 8

For any $(\mathbf{m}, \mathbf{v}) \in \mathcal{H}^*$, $\mathbf{m} = \mathbf{m}^\pi$ and from Lemma 7, $\widehat{\boldsymbol{\theta}}(t)$ converges to $\mathcal{H}^*$. Hence (a) holds.

To show (b), pick some $c \in \mathcal{C}$, and note that $\mathbf{r}^*(\boldsymbol{\theta}, c)$ is a uniformly continuous function of $\boldsymbol{\theta}$ on $\mathcal{H}$ (uniform continuity follows from the continuity of $\mathbf{r}^*(\boldsymbol{\theta}, c)$ proven in Lemma 3 (a), and compactness of $\mathcal{H}$). Hence, for any $\epsilon > 0$, we can find a $\delta > 0$ such that for any $\boldsymbol{\theta}' \in \mathcal{H}$, $d\left(\mathbf{r}^*(\boldsymbol{\theta}, c), \mathbf{r}^*(\boldsymbol{\theta}', c)\right) < \epsilon$ for any $\boldsymbol{\theta} \in \mathcal{H}$ such that $d(\boldsymbol{\theta}, \boldsymbol{\theta}') < \delta$. Here $d$ denotes the Euclidean distance metric for $\mathbb{R}^{3N}$. In particular, for any $\boldsymbol{\theta}' \in \mathcal{H}_*$, $d\left(\mathbf{r}^*(\boldsymbol{\theta}, c), \mathbf{r}^*(\boldsymbol{\theta}', c)\right) < \epsilon$ for any $\boldsymbol{\theta} \in \mathcal{H}$ such that $d(\boldsymbol{\theta}, \boldsymbol{\theta}') < \delta$. From the definition of $\mathcal{H}_*$, $\left(\mathbf{r}^*(\boldsymbol{\theta}', c)\right)_{c \in \mathcal{C}} = (\mathbf{r}^*(\boldsymbol{\theta}^\pi, c))_{c \in \mathcal{C}}$ since $\boldsymbol{\theta}' \in \mathcal{H}_*$. Thus, we have that $d(\mathbf{r}^*(\boldsymbol{\theta}, c), \mathbf{r}^*(\boldsymbol{\theta}^\pi, c)) < \epsilon$ for any $\boldsymbol{\theta} \in \mathcal{H}$ such that $d(\boldsymbol{\theta}, \boldsymbol{\theta}') < \delta$. From Lemma 7, we have that $\left(\widehat{\boldsymbol{\theta}}_t\right)_t$ converges to the set $\mathcal{H}_*$. Hence, for a sufficiently large $t$, $d\left(\widehat{\boldsymbol{\theta}}_t, \boldsymbol{\theta}'\right) < \delta$ for some $\boldsymbol{\theta}' \in \mathcal{H}^*$, and thus $d\left(\mathbf{r}^*\left(\widehat{\boldsymbol{\theta}}_t, c\right), \mathbf{r}^*(\boldsymbol{\theta}^\pi, c)\right) < \epsilon$. Thus, part (b) holds.

Parts (c) and (d) can be proved using a similar approach as above by using the following facts: (i) $\widehat{\boldsymbol{\theta}}(t)$ converges to $\mathcal{H}_*$; (ii) $\left(U_i^V\right)'(v_i) = \left(U_i^V\right)'(v_i^\pi)$ for any $(\mathbf{m}, \mathbf{v}) \in \mathcal{H}_*$; and (iii) For each $i \in \mathcal{N}$, $\left(U_i^V\right)'(.)$ is uniformly continuous on $[0, v_{\max}]$.

## Appendix E
## Proof of Lemma 9

Consider any realization $(c_t)_t$ of $(C_t)_t$. For any $c \in \mathcal{C}$, using Lemma 8 (b) and the ergodicity of $(C_t)_t$, we have

$$
\begin{aligned}
\lim_{T \to \infty} \frac{1}{T} \sum_{t=1}^T \mathbf{I}_{(c_t = c)} \mathbf{r}^*\left(\widehat{\boldsymbol{\theta}}_t, c\right) &= \mathbf{r}^*(\boldsymbol{\theta}^\pi, c) \lim_{T \to \infty} \frac{1}{T} \sum_{t=1}^T \mathbf{I}_{(c_t = c)} \\
&= \pi(c) \mathbf{r}^*(\boldsymbol{\theta}^\pi, c)
\end{aligned}
$$

Since, $\mathbf{r}^*\left(\widehat{\boldsymbol{\theta}}_t, c_t\right) = \sum_{c \in \mathcal{C}} \mathbf{I}_{(c_t = c)} \mathbf{r}^*\left(\widehat{\boldsymbol{\theta}}_t, c\right)$ and $\mathcal{C}$ is a finite set, we can use the above result to conclude that

$$
\begin{aligned}
\lim_{T \to \infty} \frac{1}{T} \sum_{t=1}^T \mathbf{r}^*\left(\widehat{\boldsymbol{\theta}}_t, c_t\right) &= \lim_{T \to \infty} \frac{1}{T} \sum_{t=1}^T \sum_{c \in \mathcal{C}} \mathbf{I}_{(c_t = c)} \mathbf{r}^*\left(\widehat{\boldsymbol{\theta}}_t, c\right) \\
&= \sum_{c \in \mathcal{C}} \lim_{T \to \infty} \frac{1}{T} \sum_{t=1}^T \mathbf{I}_{(c_t = c)} \mathbf{r}^*\left(\widehat{\boldsymbol{\theta}}_t, c\right) \\
&= \sum_{c \in \mathcal{C}} \pi(c) \mathbf{r}^*(\boldsymbol{\theta}^\pi, c) \\
&= \sum_{c \in \mathcal{C}} \pi(c) \mathbf{r}^\pi(c) \\
&= \mathbf{m}^\pi.
\end{aligned}
$$

This proves part (a).

Using the ergodicity of $(C_t)_t$, parts (b) can be proved using a similar approach (as above) by using part (c) of Lemma 8.